# The surface-topography challenge: Problem definition

Tevis D. B. Jacobs,[1] Nathaniel Miller,[1] Martin H. Müser,[2] Lars Pastewka[3]

[1] Department of Mechanical Engineering and Materials Science, University of Pittsburgh, Pittsburgh, PA 15261, US

[2] Department of Materials Science and Engineering, Saarland University, 66123 Saarbrücken, Germany

[3] Department of Microsystems Engineering, University of Freiburg, 79110 Freiburg, Germany

*Email*: tjacobs@pitt.edu, martin.mueser@mx.uni-saarland.de, lars.pastewka@imtek.uni-freiburg.de

## ABSTRACT

We present to the community a surface-definition problem, whose solution we consider to be critical for the proper description of contacts between nominally flat surfaces [1,2]. In 2015, Müser and Dapp issued the Contact Mechanics Challenge, which provided complete topography data for a fictional surface and asked theorists and modelers to compute the expected contact parameters for such a surface. This effort was a success, but exposed one glaring flaw in the community's understanding of the nature of contact: these models require as input a complete description of surface topography, which is rarely or never available for real-world surfaces [3–6]. The present challenge is to experimentalists: we will send you samples of two materials (one smoother and one rougher); you determine the surface topography of these materials. We call on you to measure such surfaces however you wish, using contact-based techniques, light scattering, microscopy, or other techniques. Examples of quantities of interest are: root-mean-square (RMS) parameters; the power spectral density (PSD); or the autocorrelation function (ACF). For the material, we have chosen chromium nitride, a wear- and corrosion-resistant coating used in industrial applications including automotive components, cutting tools, and die-casting. To participate, simply go to: https://contact.engineering/challenge to provide your shipping address and other information, then samples will be shipped out to you. The only requirement of participation is that your raw topography measurements are deposited on the free *contact.engineering* web app to facilitate data sharing. The purpose of this challenge is for our community to move towards: (a) better agreement on how to describe the multi-scale topography of experimental surfaces; and (b) better understanding of how to apply the well-developed models and theories to real-world surfaces.

## MOTIVATION FOR THE PRESENT CHALLENGE

2022 marks the 110th anniversary of Binder's realization [7] that roughness limits the area of real contact between two interfaces. Since then, the crucial importance of roughness as a factor determining almost all interfacial phenomena—such as friction, adhesion, or electrical and thermal conductivity—has permeated the physical and engineering sciences. In response, a tremendous number of investigations have experimentally measured, analyzed, described, and reported surface topography, and a similarly large volume of literature describes a wide variety of models and theories for computing functional properties from such quantitative descriptions of surface topography [1–3,8,9]. However, even after 110 years, the central challenge in designing optimal surfaces remains: how can the topography-dependent properties of rough surfaces be understood or predicted?

To advance the understanding of this question on the theoretical side, Dapp and Müser issued the *Contact Mechanics Challenge* [10], which provided complete topography data for a fictional surface and asked theorists and modelers to predict the mechanical contact properties of this surface. This effort was successful in demonstrating that many state-of-the-art numerical methods agree in central quantities (such as area and pressure distribution), but also in highlighting limitations of these methods in particular with respect to solving short-ranged adhesion. It also highlighted shortcomings of widely-used models describing surfaces as a collection of independent asperities [1]. However, taken together, the results demonstrated that a variety of different approaches can predict consistent properties.

Yet there remains a significant gap in applying these cutting-edge models to real-world surfaces, which is a lack of accurate characterization of the multi-scale topography that is required as input to these models. The surface finish of manufactured parts is often specified in terms of simple scalar quantities like $R_a$ or $R_q$ that are laid out in many competing standards [11–13] and are easy to measure, yet it is commonly known that these quantities alone cannot predict performance. Even in a scientific context, where researchers are more likely to report multi-scale metrics, e.g., the power spectral density [6] or the autocorrelation function [14], these are only measured over a limited range of scales, typically using a single technique. Even when efforts are made to measure across many scales and many techniques, the measurement approaches have experimental errors and instrumental artifacts [15–18].

This lack of knowledge of the true, multi-scale topography of real surfaces presents two distinct problems. First, simulations and numerical models cannot be adequately tested and validated with only incomplete topographic information on real surfaces. Second, the insights from these theories, models, and simulations cannot be effectively implemented on real-world surfaces. Models may predict the optimal multi-scale topography to achieve a particular surface property, yet it is not clear which surface finishing technique will get closest to this optimal topography.

Therefore the present challenge is to experimentalists: to come together as a community and to provide the most comprehensive surface description that is possible. We will send out samples of two materials to all who wish to participate. The purpose of this challenge will be to combine a large number of topography measurements into a single statistical characterization of the surface. We hope that, by combining many different instruments, using many different techniques, carried out by many different research labs, our community can learn about the accuracy, repeatability, and reproducibility of our methods of describing surface topography.

**GOALS AND OBJECTIVES**

**The overall goal of the present challenge** is for our community to move ourselves toward better understanding and agreement on how to measure, report, and analyze surface topography. This goal will be achieved through three objectives:

*Objective 1: Compare the advantages and disadvantages of different techniques for measuring surface topography.* The measurement of a single material using a wide variety of techniques and metrics enables the comparison and contrasting of results. This in turn elucidates the strengths and limitations of each technique. While some of these limitations may be obvious (e.g., an optical technique will be limited by the diffraction limit of light), for other limitations there is disagreement about the best way to measure and correct them, such as the effect of tip-induced artifacts in stylus profilometry or atomic force microscopy.

*Objective 2: Create the single most comprehensive description of a surface yet performed.* By combining all results into a single statistical description of the material's surface topography, we attempt to overcome the individual problems that are inherent to any single technique, such as instrument artifacts, noise, and limitations in scanning size or resolution. This fully comprehensive surface description will provide theoreticians and modelers with sufficient information to fully describe this surface and have a publicly available real-world dataset that can be used as an input to any calculation.

*Objective 3: Aid the development of next-generation surface descriptors.* Most of the current scale-dependent statistical descriptions of surface topography use simplifying assumptions. For example, the distributions of surface height or surface slope are often approximated as Gaussian. By collecting and publishing the raw topography data for an extremely well-characterized surface, this challenge enables the evaluation of the accuracy of these assumptions and may facilitate the generation of wholly new descriptors that more accurately describe the topography.

## **LOGISTICS**

*How to participate*: Please visit https://contact.engineering/challenge. The basic process is as follows:

1. Enter your information at: https://contact.engineering/challenge
2. Samples will be shipped out to you. Each sample is labeled with a unique ID.
    a. (Sample shipment will begin in August, 2022; and continue until the conclusion of the challenge).
3. Please measure and analyze the surface topography, using any method(s) you like
4. Please upload your raw topography data onto *contact.engineering*
    a. See section below on *How to share topography data*
    b. *NOTE: Please DO NOT publish your data yet. Data should later be published (using the publish function in *contact.engineering*, which includes DOI) after the challenge is concluded, but we don't want "spoilers" ahead of time!*
5. All participants will be co-authors on a concluding publication for this challenge, which reports results in *Tribology Letters* (subject to the journal's normal reviewing procedures).
6. OPTIONAL: Of course if there is something novel in your methodology, or you wish to measure other properties of these samples, you are free to independently submit a separate manuscript about your findings.

*The material of interest* is chromium nitride (CrN). Two different types of samples will be sent:

1. A "smoother surface" – CrN deposited on a prime-grade polished silicon wafer;
2. A "rougher surface" – CrN deposited on the rough "backside" of a single-side-polished silicon wafer, which has been subsequently etched with isotropic reactive ion etching.

CrN was chosen because it is a wear- and corrosion-resistant coating that is widely used in automotive components, cutting tools, and die-casting [19–21]. CrN is typically deposited via physical vapor deposition (PVD), and the present deposition uses a magnetron sputtering technique. The silicon substrates were chosen due to their extreme reproducibility in fabrication. The two substrates are intended to produce a "smoother surface" that is representative of materials used in the semiconductor industry, and a "rougher surface" that has larger topographic variation, as is common in other industrial contexts. The standard sample size is 1 cm x 1 cm, shipped in a standard wafer box; although custom sizes can be provided upon request.

*How to measure topography*: You are encouraged to use any and all techniques at your disposal. These can include, but are not limited to, stylus profilometry (line and/or area scans), optical interferometry, 3D microscopies (e.g., scanning focus variation or confocal), structured illumination, stereo- or angle-resolved microscopy, cross-section or side-view microscopy, x-ray reflectivity, small-angle x-ray scattering (SAXS), small-angle neutron scattering (SANS), or any of the scanning probe microscopy/metrology techniques.

*How to analyze your topography data:* You are free to analyze your topography using any and all methods that you choose. We recognize that different fields have different preferred metrics, where some contexts prefer simple scalar metrics such as $R_a$ and $R_q$, and other contexts prefer scale-dependent metrics like the PSD or the ACF. Likewise, there are different conventions about methods for computing those metrics, with some fields using the built-in software on characterization machines, and other fields using freely available tools like Gwyddion, while still other fields prefer to typically write their own code in MATLAB or Python. (You are welcome to use the open-source analysis routines available at *contact.engineering*, but you are not required to do so!) Part of the purpose of this challenge is to identify the points of commonality and points of diversion between different conventions – for that reason, we encourage you to report the results as you typically would in a scientific publication in your field.

*How to share the raw topography data:* In order to share raw data in a mutually accessible form, please upload your raw topography data to the freely available web-app available at https://contact.engineering [22]. Please contact us if your instrument's native file format is not yet supported. By default, all uploads to the site are securely stored and private, until you choose to share with collaborators or publish the data (which gives it a DOI). Note that *contact.engineering* is designed so that you can combine many measurements (including from disparate techniques) into a single *digital surface twin*. Of course you may create as many different private digital surface twins as you like while you are taking and analyzing data.

When you are ready to submit, we request that you create one digital surface twin for each sample that you measured, and upload all raw topography data, from all techniques, into that container. *One digital surface twin per sample*; and please include the sample ID in the name of your digital surface twin. ***NOTE: Please DO NOT publish your data (yet); instead, please use *contact.engineering* to share it with the four authors of this document. At the conclusion of the challenge, we will ask that you publish all of your data.*** Your data will be combined with that of all other participants to create comprehensive statistical descriptions of the surface topography of these materials.

*How to report methods and results:* Please submit a concise discussion of the methodology for all measurements performed. The description can be informal and it will be our duty to combine all text into a coherent description, in which similar methods from different participating groups are grouped together to avoid redundancy.

Please also submit a concise description of your results, including which metrics and parameters you computed, and how the calculations were performed. If possible, we encourage you to specify the number of significant figures and also to estimate the uncertainty in your measured values. Please report the results in a data file (.asc, .xls, etc), with clear descriptions of all columns and rows (including variable names and specifying units).

*Optional:* Should your submission contain novel methodology, you are (of course) free to submit a companion or follow-up paper to the surface topography challenge. Likewise, if you choose to measure some functional property of these surfaces, you are free to report that in a separate publication.

*Authorship:* Please include only the authors that directly contributed (or supervised) the work. We will ask for author contribution statements if your group participates with more than four authors.

*Submission deadline:* **August 31, 2023.**

We encourage the surface-science, tribology, and surface-metrology communities to participate in this challenge. Together we can drive forward the state of the art, and improve the reliability and interpretability of surface topography measurements.